\documentstyle[12pt]{article}
\begin{document}
\begin{center}
\textbf{THE  CASIMIR  PROBLEM  OF  SPHERICAL  DIELECTRICS: NUMERICAL  EVALUATION  FOR  GENERAL  PERMITTIVITIES}\\
\bigskip
\bigskip

 I. Brevik\footnote{Email address: iver.h.brevik@mtf.ntnu.no} and J. B. Aarseth\footnote{Email address: jan.b.aarseth@mtf.ntnu.no} \\

\bigskip
\bigskip

Division of Applied Mechanics, Norwegian University of Science and Technology,\\
N-7491 Trondheim, Norway\\

\bigskip
\bigskip

J. S. H{\o}ye\footnote{Email address:  johan.hoye@phys.ntnu.no}\\

\bigskip
\bigskip

Department of Physics, Norwegian University of Science and Technology,\\
N-7491 Trondheim, Norway\\

\bigskip
\bigskip
PACS numbers:  05.30.-d; 05.40.+j; 34.20.Gj; 03.70.+k\\
\bigskip
 Revised version, June 2002\\
\end{center}
\bigskip
\bigskip

\begin{abstract}

The Casimir mutual free energy $F$ for a system of two dielectric  concentric nonmagnetic spherical bodies is calculated, at arbitrary temperatures. The present paper is a continuation of an earlier investigation [Phys. Rev. E {\bf 63}, 051101 (2001)], in which $F$ was evaluated in full only for the case of ideal metals (refractive index $n=\infty$). Here, analogous results are presented for dielectrics, for some chosen values of $n$. Our basic calculational method stems from quantum statistical mechanics. The Debye expansions for the Riccati-Bessel functions when carried out to a high order are found to be very useful in practice (thereby overflow/underflow problems are easily avoided), and also to give accurate results even for the lowest values of $l$ down to $l=1$. Another virtue of the Debye expansions is that the limiting case of metals becomes quite amenable to an analytical treatment in spherical geometry. We first discuss the zero-frequency TE mode problem from a mathematical viewpoint and then, as a physical input, invoke the actual dispersion relations. The result of our analysis, based upon the adoption of the Drude dispersion relation at low frequencies, is that the zero-frequency TE mode does not contribute for a real metal. Accordingly, $F$ turns out in this case to be only one half of the conventional value at high temperatures. The applicability of the Drude model in this context has however been questioned recently, and we do not aim at a complete discussion of this issue here. Existing experiments are low-temperature experiments, and are so far not accurate enough to distinguish between the different predictions. We also calculate explicitly the contribution from the zero-frequency mode for a dielectric. For a dielectric, this zero-frequency problem is absent.

\end{abstract}

\newpage

\section{Introduction}

In the Casimir world, it is desirable to consider geometrical configurations that are amenable to an analytical treatment and at the same time nontrivial enough to elucidate the physically important properties. The following configuration turns out to satisfy these two criteria (cf. Fig.~1): there are two spherical bodies present with concentric surfaces at $r=a$ and $r=b$, with a vacuum region in between. We shall consider the free energy $F(T)$ due to the mutual interaction between the two bodies. We gave earlier an analysis of this problem \cite{hoye01}, with the use of quantum statistical methods as well as field theoretical methods. Whereas the general formalism in \cite{hoye01} was valid for arbitrary (equal) permittivities $\varepsilon$ in the two dielectric regions $r<a$ and $r>b$, the explicit evaluation of $F(T)$ for various values of $T$ and widths $d=(b-a)$ was made for the case of {\it perfectly conducting walls} only, corresponding to $\varepsilon \rightarrow \infty$. Our purpose with the present paper is to extend these previous considerations to cover the case of general values of the permittivity. As to our knowledge such a calculation has not been undertaken before, although there are similarities with the theory given by Kleinert some years ago \cite{kleinert89}. We will henceforth assume, as in \cite{hoye01}, that the two media are nonmagnetic. A brief account of the essentials of the present theory was recently given in Ref.~\cite{brevik02}. We will have to repeat some of the formalism below because of readability.

One lesson from the calculation in \cite{hoye01} was that the power of the 
 quantum statistical method is remarkably strong when applied to the rather demanding case of general $\varepsilon$. The most central formula in our context is the statistically-derived Eq.~(40) in \cite{hoye01}; it gives the value of $\beta F \equiv F/T$ for arbitrary values of temperature, width, and $\varepsilon$. Whereas this equation was given in terms of a very compact notation in \cite{hoye01}, it will be convenient here to rewrite it slightly. Let $m \in \langle -\infty, \infty \rangle$ be an integer corresponding to Matsubara frequencies $K=2\pi m/\beta$; let $n=\sqrt \varepsilon$ be the refractive index of the two media lying at $r<a$ and $r>b$, and let  $s_l(x),~e_l(x)$ be Riccati-Bessel functions with imaginary argument defined according to $s_l(x)=(\pi x/2)^{1/2}I_\nu(x),~~e_l(x)=(2x/\pi)^{1/2}K_\nu(x)$,
so that their Wronskian becomes $W\{s_l, e_l \}=-1$. Here $\nu=l+1/2$, and $I_\nu,~K_\nu$ are modified Bessel functions. We write the formula as
\begin{equation}
\beta F={\sum_{m=0}^\infty}' \sum_{l=1}^\infty (2l+1)[\ln (1-\lambda_l^{TM})+
\ln(1-\lambda_l^{TE}) ],
\end{equation}
\label{1}
where the prime on the summation sign means that the $m=0$ term is taken with half weight. The two eigenvalues $\lambda_l^{TM}$ and $\lambda_l^{TE}$ in Eq.~(1) correspond to the transverse magnetic and the transverse electric modes. (In the notation of Ref.~\cite{hoye01},  $~ \lambda_{\varepsilon l} \equiv \lambda_l^{TM},~\lambda_l \equiv \lambda_l^{TE}$; we find it here useful to emphasize the physical nature of the two modes.) For later use we will write these eigenvalues as ratios. First,
\begin{equation}
\lambda_l^{TM}=\frac{f_1 f_2}{f_3 f_4},
\end{equation}
\label{2}
where
\[ f_1= ns_l'(x)s_l(nx)-s_l(x)s_l'(nx) ,\]
\[ f_2= ne_l'(y)e_l(ny)-e_l(y)e_l'(ny), \]
\[ f_3= ne_l'(x)s_l(nx)-e_l(x)s_l'(nx), \]
\begin{equation}
f_4= ne_l(ny)s_l'(y)-e_l'(ny)s_l(y),
\end{equation}
\label{3}
$x$ and $y$ being the nondimensional frequencies
\begin{equation}
x=2\pi ma/\beta,~~~~y=2\pi mb/\beta.
\end{equation}
\label{4}
We put $\hbar =c=k_B =1;~\beta = 1/T$ is the inverse temperature. It should be noted that, in contradistinction to the formalism in \cite{hoye01}, the primes in Eqs.~(3) mean derivatives with respect to the {\it whole} argument. 

Next, the TE eigenvalues are written as
\begin{equation}
\lambda_l^{TE}=\frac{g_1 g_2}{g_3 g_4},
\end{equation}
\label{5}
where
\[ g_1=s_l'(x)s_l(nx)-ns_l(x)s_l'(nx), \]
\[ g_2= e_l'(y)e_l(ny)-ne_l(y)e_l'(ny), \]
\[ g_3= e_l'(x)s_l(nx)-ne_l(x)s_l'(nx), \]
\begin{equation}
 g_4= e_l(ny)s_l'(y)-ne_l'(ny)s_l(y).
\end{equation}
\label{6}
The case of metallic walls, $n=\sqrt{\varepsilon} \rightarrow \infty$, leads to a delicate two-limit problem as regards the contribution from zero Matsubara frequency, $m=0$.  The conventional way to proceed when handling this problem within the framework of nondispersive theory, has been to take the limits in the following order:

(i)  Set first $\varepsilon = \infty$;

(ii)  take then the limit $m\rightarrow 0$. 

This way of taking the limits was advocated already in the 1978 paper of Schwinger, DeRaad, and Milton \cite{schwinger78}, and the same procedure was followed in Sec.~VII of our previous paper \cite{hoye01}. The method implies inserting the small-argument approximations for the Riccati-Bessel functions into Eq.~(1), resulting in the following free energy expression:
\begin{equation}
\beta F(\varepsilon \rightarrow \infty)={\sum_{m=0}^{\infty}}'\sum_{l=1}^{\infty}
(2l+1)\ln \left\{ \left[ 1-\frac{s_l(x)}{e_l(x)}\frac{e_l(y)}{s_l(y)}\right]
\left[ 1-\frac{s_l'(x)}{e_l'(x)}\frac{e_l'(y)}{s_l'(y)}\right] \right\},
\end{equation}
\label{7}
which is in agreement with Eq.~(68) in \cite{hoye01}. If we next let $x\rightarrow 0,~y\rightarrow 0$ observing the same low-argument approximations, we obtain as contribution from $m=0$:
\begin{equation}
\beta F^{conv}(\varepsilon \rightarrow \infty, m=0)=\sum_{l=1}^\infty (2l+1)\ln \left[ 1-\left( \frac{a}{b}\right)^{2l+1} \right],
\end{equation}
\label{8}
again in agreement with \cite{hoye01}, Eq.~(79). This is the conventional result. {\it Both} the two electromagnetic modes are in this way found to contribute equally to the sum in Eq.~(8).

However, a discussion has recently arisen as to whether this recipe for dealing with the $m=0$ term in the TE mode is really correct. The problem becomes most acute in the high $T$ regime, but is present at moderate and low temperatures also. We may refer to the paper of Bostr\"{o}m and Sernelius \cite{bostrom00} questioning this point, and the subsequent comment of Lamoreaux \cite{lamoreaux01}. What has been most welcome in recent years are the accurate experiments on the Casimir force, due to  Lamoreaux \cite{lamoreaux97}, Mohideen {\it et al.} \cite{mohideen98, roy99, harris00, chen02}, and Bressi {\it et al.}\cite{bressi02}. By means of these experiments it becomes much easier to formulate a sound theory. Several theoretical papers have appeared, discussing various facets of the experiments \cite{lamoreaux98, lambrecht00, svetovoy00, bordag00, klimchitskaya01}. An extensive recent review has been given by Bordag {\it et al.} \cite{bordag01}. We also mention several other related papers \cite{milton99, barton01, feinberg01}, of a more general nature, though also these are being concerned with finite temperature effects in a Casimir context.

One of the purposes of the present work is to analyze how the $m=0$ case works out for case of the spherical geometry of Fig.~1. It turns out that the formalism becomes actually quite manageable. Use of the Debye expansion for the Riccati-Bessel functions is an essential element in our analysis, and it implies as an additional bonus that the overflow and underflow problems that so often plague calculations of this sort, are easily abandoned.

Our analysis obviously requires machine evaluation, as we will carry out the Debye expansion to the 18th order in the quantity $\theta$ defined in Sec. 2.1 below. Then the numerical accuracy becomes quite satisfactory for all practical purposes, even for the lowest values of $l$ down to $l=1$. Up to about one million terms in the series will be summed. The limiting case of metals will be handled in a physical rather than a mathematical way by adopting the physically preferable Drude dispersive model as input at low frequencies. On basis of the Drude dispersion relation, we are quite naturally led to the conclusion that there is no contribution from the $m=0$ TE mode to the free energy $F$ in the case of infinite conductivity. This implies that the conventional expression for $F$ for metals has to be multiplied by one half. (This conclusion is actually in agreement with the outcome of the statistical-mechanical considerations in Sec.~III in \cite{hoye01}.)  We show graphically several results for how $F$ varies with temperature and width, both for the ideal metallic case (in which the zero-frequency mode is counted twice) and for the dielectric case. Finally, we calculate the magnitude of the $m=0$ contribution to $F$ for a dielectric, as a function of temperature, and compare the result with the total value of $F$. Also, the mutual internal energy $E$ itself is briefly discussed.

We do not in the present paper aim at resolving the issue about the $m=0$ term for the TE mode for real metals. We make however some estimates in the discussion in Sec. 5, item 6, choosing aluminum as a concrete example. As for experiments, the atomic force microscope experiment of Mohideen and Roy \cite{mohideen98}, and that of Harris {\it et al.} \cite{harris00}, achieved an accurary of about one per cent. These are essentially low-temperature experiments, where the influence from the $ m=0$ TE term is small (at $T=0$ the $m=0$ is completely negligible since the sum over discrete Matsubara frequencies is replaced by an integral over imaginary frequencies). The all-over temperature correction at room temperature is predicted to be of the same one per cent accuracy \cite{klimchitskaya01} . The single $m=0$ temperature term which we discuss, is not singled out experimentally under these circumstances.

\section{Numerical considerations}

We now define the nondimensional temperature:
\begin{equation}
t=\frac{2\pi a}{\beta},
\end{equation}
\label{9}
implying that $x=mt$. 
It turns out numerically that the conventional uniform asymptotic expansions of the Riccati-Bessel functions, which are used often to low orders when dealing with rough approximations, become quite accurate if the polynomial parts of the expansions are expanded to high order. This makes the evaluation in the present case straightforward in principle: the polynomial parts, which generally turn out to be about unity in magnitude, can easily be handled on a computer. The remaining parts of the Bessel functions, which are simple exponentials, can be dealt with analytically. Actually, what are needed in practical calculations, are {\it fractions} between Bessel functions. 
This is the way in which the overflow or underflow problems are avoided. Overflow/underflow problems would easily arise if we instead chose to take the whole Bessel function directly from the computer library. The remaining numerical evaluation is not quite trivial, though; especially in the case of a narrow slit the number of necessary terms turns out to be quite large, about $10^6$, as we mentioned above.

We start by presenting our expanded version of the Debye formalism.

\subsection{The Debye expansions}

Let us write the Debye expansions in the form 
\begin{equation}
s_l(x)=\frac{1}{2}\frac{\sqrt{z(x)}}{[1+z^2(x)]^{1/4}}\, e^{\nu \eta(x)}\,A[\theta(x)],
\end{equation}
\label{10}
\begin{equation}
e_l(x)=\frac{\sqrt{z(x)}}{[1+z^2(x)]^{1/4}}\,e^{-\nu \eta(x)}\,B[\theta(x)],
\end{equation}
\label{11}
\begin{equation}
s_l'(x)=\frac{1}{2}\frac{[1+z^2(x)]^{1/4}}{\sqrt{z(x)}}\,e^{\nu \eta(x)}\,C[\theta(x)],
\end{equation}
\label{12}
\begin{equation}
e_l'(x)=-\frac{[1+z^2(x)]^{1/4}}{\sqrt{z(x)}}\,e^{-\nu \eta(x)}\,D[\theta(x)].
\end{equation}
\label{13}
Here $\nu=l+1/2,~l=1,2,..., \quad 
 z(x)=x/\nu,~~~~\theta(x)=[1+z^2(x)]^{-1/2}, $ and
\begin{equation}
\eta(x)=\frac{1}{\theta(x)}+\ln \frac{z(x)}{1+1/\theta(x)} 
\end{equation}
\label{14}
($\theta$ is the same as the symbol $t$ in Ref.~\cite{abramowitz72}). There occur four polynomials, $A(\theta), B(\theta), C(\theta), D(\theta)$, which are found to be of order unity. In Ref.~\cite{abramowitz72} the first two of them, $A(\theta)$ and $B(\theta)$, are expanded to order $\theta^{12}$, whereas $C(\theta)$ and $D(\theta)$ are expanded to order $\theta^9$. In Ref.~\cite{brevik87} we expanded all the polynomials to order $\theta^{18}$. These expansions, which will not be reproduced here, are found to be easily handled on a computer. The polynomials possess the important property that they go to unity when $\theta$ goes to zero.

The factors in Eqs.~(10)-(13) that can take extreme values, are the exponentials. They are easily dealt with analytically.

It is now convenient to calculate the following ratios between the functions defined in Eqs.~(3):
\begin{equation}
\frac{f_1}{f_3}=-\frac{1}{2}e^{2\nu \eta(x)}\,\frac{n^2\gamma C[\theta(x)]-A[\theta(x)]C[\theta(nx)]/A[\theta(nx)]}
{n^2\gamma D[\theta(x)]+B[\theta(x)]C[\theta(nx)]/A[\theta(nx)]},
 \end{equation}
\label{15}
\begin{equation}
\frac{f_2}{f_4}=-2e^{-2\nu \eta(y)}\, \frac{n^2\delta D[\theta(y)]-B[\theta(y)]D[\theta(ny)]/B[\theta(ny)]}
{n^2\delta C[\theta(y)]+A[\theta(y)]D[\theta(ny)]/B[\theta(ny)]},
\end{equation}
\label{16}
where $\gamma$ and $\delta$ are the coefficients
\begin{equation}
\gamma= \sqrt{\frac{1+z^2(x)}{1+z^2(nx)}},~~~~\delta=\sqrt{\frac{1+z^2(y)}{1+z^2(ny)}}.
\end{equation}
\label{17}
Similarly
\begin{equation}
\frac{g_1}{g_3}=-\frac{1}{2}e^{-2\nu \eta(x)}\,\frac{\gamma C[\theta(x)]-A[\theta(x)]C[\theta(nx)]/A[\theta(nx)]}
{\gamma D[\theta(x)]+B[\theta(x)]C[\theta(nx)]/A[\theta(nx)]},
\end{equation}
\label{18}
\begin{equation}
\frac{g_2}{g_4}=-2e^{-2\nu \eta(y)}\, \frac{\delta D[\theta(y)]-B[\theta(y)]D[\theta(ny)]/B[\theta(ny)]}
{\delta C[\theta(y)]+A[\theta(y)]D[\theta(ny)]/B[\theta(ny)]}.
\end{equation}
\label{19}
Now the eigenvalues $\lambda_l^{TM}$ and $\lambda_l^{TE}$ are calculable from Eqs.~(2) and (5), with use of the expressions (15)-(19) in which the $\theta$-expansions for the four polynomials are taken from Ref.~\cite{brevik87}.
As in our previous paper \cite{hoye01}, we made use of standard FORTRAN routines throughout.

\subsection{Calculated results for dielectrics}

On a logarithmic plot with base 10, Fig.~2 shows how $\log_{10}(-\beta Ft)$ varies with $d/a$ for various values of $t$ when the medium is dilute, $n=1.1$. The figure is to be compared with the corresponding Fig.~1 in \cite{hoye01}. As expected, the magnitude $|F|$ of the mutual free energy is much less for a dilute medium than it is in the case of ideal metallic walls ($n=\infty$). For instance, when $d/a=0.2,~t=1$, for $n=1.1$ we see that $|F|$ has only about 0.1 \% of the value it has for an ideal  metal. The various curves in Fig.~2 tend to overlap at low temperatures. Thus the curve calculated for $t=0$ turns out to be indistinguishable from the curve calculated for low temperatures up to $t=1$. The curves in Fig.~2 are most useful for the case of low temperatures.

Figure 3 shows how $\log_{10}(-\beta F)$ varies with $d/a$. This representation is convenient for the case of high temperatures, since the curves for high $t$ tend to overlap.

Figure~4 shows the representation in the form that is probably the most instructive one, namely $\log_{10}(-\beta Ft)$ as a function of $\log_{10}t$. It shows clearly how there is a low-temperature plateau, extending up to a region lying somewhere between 1 and 2 in the cases shown. For higher values of $t$, there is a gradual change into the region where $F$ varies linearly with $t$.

We  calculated analogous figures for other values of $n$ also, with results as one would expect: the influence of the medium becomes strengthened when $n$ becomes greater. Figure 5 shows, as an example,  the analogue of Fig.~4 in the case of $n=2$. For instance, when $t=1,~d/a=0.01$, the magnitude $|F|$ is about 50 times as large when $n=2$ as when $n=1.1$.

Figure~6 shows the analogous variation of $F$ for the case of an ideal metal (i.e., $n=\infty$ for all $\hat{\omega}$, the $m=0$ mode counted twice). This figure is reproduced from Fig.~3 in \cite{hoye01}; it is included here both for the purpose of comparison, and also to correct the labeling on two of the curves in our previous Fig.~3. We remind ourselves that the order of taking the limits in \cite{hoye01} was in accordance with the prescriptions (i) and (ii) mentioned earlier, above Eq.~(7) in Sec.~1 (Refs.~\cite{schwinger78,hoye01}).

Generally, we found the asymptotic Debye expansions to be useful for $x>10$ and/or $l>9$. Then, an accuracy of 8 digits for the individual terms was achieved. Below these limits for $x$ and $l$, we employed the machine-generated Bessel functions. For small values of $d/a$ and $t$, slow convergence was observed. The summation of the series thus became rather demanding.  For instance, when $d/a=0.05,~t=0.01$ about 1.1 million terms were needed, if we truncated the summation at  $\varepsilon =10^{-9}$ (here $\varepsilon$ means the ratio between a general term in the series and the sum). The sum itself is however accurate only up to 4 or 5 digits.

An important result was that even for low values of $l$, the asymptotic series gave very good results. One reason for this is the high-order expansions used for the polynomials $A,B,C,D$. Most probably, the Debye expansions (at least when carried out to order $\theta^{18}$) can be used for {\it all} $x$ and $l$, for all practical purposes.

\section{The limiting case of a metal}

\subsection{The nondispersive case}

Although it would seem most natural to discuss the case of a metal on the basis of a parallel-plates configuration, let us analyze here how the formalism behaves in the ideal-conductor limit when the spherical geometry of Fig.~1 is given. It will actually turn out that the Debye expansion is very useful also in this case. As we treated this topic in reasonable detail in Ref.~\cite{brevik02}, we need only to be brief here.

We assume first that the medium is nondispersive. The formal limit that we have to take, is thus $\varepsilon \rightarrow \infty$. Let us categorize how to take the two actual limits: we let option A mean {\it first} taking the refractive index $n= \sqrt{\varepsilon}\rightarrow \infty$, {\it thereafter} taking the Matsubara frequency $m \rightarrow 0$.  Option B reverses the succession of the limits on $n$ and $m$. 

Consider first the TM mode, employing option A. When $n\rightarrow \infty$, $\theta(x)$ is finite, while $\theta(nx)\rightarrow 0$. Thus all polynomials $\{A,B,C,D\}[\theta(x)]$ at argument $\theta(x)$ are finite, while $\{A,B,C,D\}[\theta(nx)]\rightarrow 1$. Observing that $n^2\gamma$ and $n^2\delta$ are proportional to $n$ for large $n$ we get, when taking the limit $m\rightarrow 0$, the following expression for the $m=0$ contribution to the TM free energy:
\begin{equation}
\beta F^{TM}(m=0)=\frac{1}{2}\sum_{l=1}^\infty (2l+1)\ln \left[1-\left(\frac{a}{b}\right)^{2l+1} \right].
\end{equation}
\label{20}
Following instead option B we find precisely the same expression as in Eq.~(20). The $m=0$ TM free energy is thus robust with respect to the choice between options A and B. This is actually what we would expect physically: the TM mode means that the magnetic field is transverse to the radius vector ${\bf r}$ at $r=a,b$; this is the natural electromagnetic boundary condition at perfectly conducting surfaces.

Consider then the TE mode, employing option A. The difference from the preceding case lies in the sensitivity of $\lambda_l^{TE}(n\rightarrow \infty)$ with respect to the coefficients $\gamma$ and $\delta$. From Eq.~(17) we get $\gamma \rightarrow 0,\, \delta \rightarrow 0$ implying that, in the limit $m\rightarrow 0$, $\lambda_l^{TE}\rightarrow (a/b)^{2l+1}$. It follows that the TE contribution to the $m=0$ free energy is the same as  given by Eq.~(20).

Employing instead option B we obtain $\gamma \rightarrow 1,\, \delta \rightarrow 1$, resulting in $\lambda_l^{TE}\rightarrow 0$ when $m\rightarrow 0$. Consequently
\begin{equation}
B:\quad \beta F^{TE}(m=0)=0.
\end{equation}
\label{21}
Option B gives accordingly one half of the conventional result of Eq.~(8) for the total free energy in the ideal conductor limit.

The immediate question is now: which of the two options is correct?  We cannot decide upon this only by investigating  how the mathematical formalism behaves in the limiting cases; we have to bring physics into the consideration. That means, we have to consider a physically appropriate dispersion relation in the limit of low frequencies. That is the topic of the next subsection.

\subsection{The dispersive case}

Let now $\hat{\omega}$ denote the frequency along the imaginary frequency axis.There are mainly two actual dispersion relations on the market. The first corresponds to the {\it plasma model} of the dielectric:
\begin{equation}
\varepsilon(i\hat{\omega})=1+\frac{\omega_p^2}{\hat{\omega}^2},
\end{equation}
\label{22}
$\omega_p$ being the plasma frequency. As mentioned by Landau and Lifshitz (Sec. 78 in \cite{landau84}), the range of frequencies over which this formula is applicable begins, in practice, at the far ultraviolet for light elements and at the X-ray region for heavier elements. Let us for convenience rewrite the coefficients (17) as
\begin{equation}
\gamma= \sqrt{\frac{1+(\hat{\omega}a/\nu)^2}{1+(n\hat{\omega}a/\nu)^2}},
~~~\delta=\sqrt{\frac{1+(\hat{\omega}b/\nu)^2}{1+(n\hat{\omega}b/\nu)^2}}.
\end{equation}
\label{23}
When $\hat{\omega}\rightarrow 0$, it follows from Eq.~(22) that $n(i\hat{\omega})\hat{\omega}\rightarrow \omega_p$, which means that $n(i\hat{\omega})\hat{\omega}a/\nu \rightarrow x_p/\nu$ where, in dimensional units, $x_p \equiv \omega_p a/c$. Taking, for instance,  $\omega_p \sim 3\times 10^{16}~{\rm s}^{-1}$ and $a \sim 1$ cm we get $x_p \sim 10^6$. In practice, the most significant values of $l$ are much lower than this.  We can thus assume that $x_p/\nu \gg 1$ in Eq.~(23), so that in practice $\gamma \rightarrow 0,~\delta \rightarrow 0$. That is, we recover in this way {\it option} A. In conclusion, the use of the plasma dispersion relation (38), to a good approximation, leads to the conventional result (8) for the $m=0$ total free energy for a metal.

Consider next the {\it Drude model} for the dielectric, corresponding to
\begin{equation}
\varepsilon(i\hat{\omega})=1+\frac{\omega_p^2}{\hat{\omega}(\hat{\omega}+\gamma)},
\end{equation}
\label{24}
$\gamma$ being the relaxation frequency. According to this relation $n(i\hat{\omega})\hat{\omega} \rightarrow 0$ when $\hat{\omega}\rightarrow 0$, implying that $\gamma \rightarrow 1,~\delta \rightarrow 1$ according to Eq.~(23). That is, we recover {\it option} B. The total $m=0$ free energy for a metal is thus according to the Drude model predicted to be one half of the expression (12).

When deciding between these dispersion relations, we expect that relation (24) is physically correct in the limit when $\hat{\omega}\rightarrow 0$.  On general grounds the permittivity has to be inversely proportional to the frequency at low frequencies; cf. Sec. 77 in \cite{landau84}. Explicitly, $\varepsilon(\omega)\rightarrow i\sigma/\omega$, or $\varepsilon(i\hat{\omega})=\sigma/\hat{\omega}$, where $\sigma$ is the conductivity. This is a result following directly from Maxwell's equations. The Drude model satifies this requirement. Thus both the Drude model (and, as we have seen, statistical mechanical methods), support the option B above. The plasma model, Eq.(22), as we have noted, is appropriate only for the higher frequencies.

\section{Calculation of the m=0 contribution to F for a dielectric}

The delicate $m=0$ problem in the limiting case of a metal accentuates the following question: how large is the $m=0$ contribution to the free energy in the general case, for a dielectric? As the last point in our paper we shall calculate this effect, for a given value of $n$, and show the result graphically in a typical example. This point appears to be of physical interest, and with the above formalism the calculation can be easily effectuated. We now return to nondispersive theory again, so that $n$ is taken to be a constant. Since the $m=0$ case does not contribute to the TE mode at all, for any finite value of $n$, our present discussion has no bearing on the topic discussed in the previous section.

Let us first summarize, from a physical point of view, how the free energy is distributed over the various frequencies for various values of the temperature. When $T\rightarrow 0$ the Matsubara frequencies are closely spaced (at $T=0$ the Matsubara summation being replaceable by an integral), and a large number of eigenfrequencies contributes to $F$. The contribution from the lowest term $m=0$ is insignificant. When $T$ increases, the number of contributing Matsubara terms gradually becomes smaller and the frequencies gathered at the lower end of the spectrum until finally, at $T\rightarrow \infty$, the term $m=0$ dominates completely (this is the classical limit). How this gradual change actually occurs, as a function of $T$, for a given relative slit width $d/a$, can however only by found by an explicit calculation.

We recall that for a given geometry there are still three quantities to be contemplated, namely $\{n, m, t\}$. Let us fix the value of $n$, and look for the contribution to $F$ from $m=0$, as a function of $t$. From Eq.~(1) we have, for an arbitrary temperature,
\begin{equation}
\beta F(m=0)=\frac{1}{2}\sum_{l=1}^\infty (2l+1)\ln (1-\lambda_l^{TM})
\end{equation}
\label{25}
(as noted, $\lambda_l^{TE}$  does not contribute for a dielectric). We define $Y$ as the ratio between  $F(m=0)$ and the expression (1) for the full free energy:
\begin{equation}
Y=\frac{F(m=0)}{F}.
\end{equation}
\label{26} 
For given $d/a$, $Y$ thus becomes a function only of $t$. Figure~7 shows $Y$ versus $t$ for various values of $d/a$, for the case when $n=1.1.$ The curves behave qualitatively as we would expect; the contribution from $m=0$ goes to zero at very low temperatures, and goes towards unity at high $t$. Analogous curves for other values of $n$ behave similarly; thus the curves calculated for $n=2.0$ turn out to be practically indistinguishable from those in Fig.~7.

One additional conclusion to be drawn from Fig.~7 is that the less the value of $d/a$, the less becomes the importance of the $m=0$ term. It is worth noticing that this is a result that can be understood physically: when the slit is narrow, as assumed in Fig.~7, we can approximately regard the system as a conventional two-plates system. For the latter geometry, it is  known that the classicality condition can be written as $dT \gg 1$, where $d$ is the distance between the plates (cf. Sec. 82 in \cite{landau80}). When $d$ decreases the system thus becomes more and more a quantum-mechanical system, necessitating an increasing large region of frequencies determining the value of $F$. The relative importance of the low frequencies, in particular that of $m=0$, thus has to diminish, in accordance with the figure.

\section{Conclusive remarks}

Let us summarize our work, and supply the above with some further remarks.

\bigskip

1.  The Debye expansion procedure is almost surprisingly effective. When carried out to sufficiently high order in the parameter $\theta$ - order 18 in the present paper - the accuracy becomes fully satisfactory for all practical purposes for all values of $l$, even down to the lowest value $l=1$. Moreover the formalism becomes straightforward to analyze, even in the delicate two-limit case $n\rightarrow \infty,~m\rightarrow 0$ associated with a  metal. As mentioned at the beginning of  Sec. 3.1  that it would seem most natural to analyze the limit of a metal assuming the standard Casimir configuration of parallel plates. In some sense the situation seems in fact to be the reverse: the spherical geometry is more easy to analyze in the metallic limit than the planar one. The reason for this is obvious: once plane plates are involved, one becomes confronted with two infinite spatial dimensions (the linear extensions of the plates), which lead to mathematically more delicate issues. Recent investigations of the Casimir effect for perfectly conducting plates have been given by Klimchitskaya, Mostepanenko, and Geyer \cite{klimchitskaya01, geyer01}. 

2.  The basic expression for the free energy $F$, Eq.~(1), holds for arbitrary temperatures as well as for arbitrary (frequency) dispersion relations. In the special case of a real metal we find, when adopting the Drude relation (in our opinion the preferable one at low frequencies)  that the $m=0$ TE mode does not contribute. The total $m=0$ free energy for a metal becomes accordingly one half of the conventional expression (8).

3.  For a dielectric (finite $n$) there is no limiting problem: the $m=0$ case does not contribute to the TE mode at all. Figure 7 shows the magnitude of the $m=0$ free energy (thus associated with the TM mode) relative to the full free energy $F$. The relative contribution from the $m=0$ term is seen to increase with temperature, as one would expect physically; the relative weight of the low Matsubara frequencies becoming enhanced at high $T$.

4.  It ought to be stressed that $F$ means everywhere the {\it mutual} free energy between the two concentric dielectric bodies. Thus $F\rightarrow 0$ when $d =(b-a)\rightarrow \infty$. It may be of interest to calculate the mutual internal energy $E$ also. By means of the thermodynamic relation $E=\partial (\beta F)/\partial \beta$ we find immediately from Eq.~(1)
\begin{equation}
E=-{\sum_{m=0}^\infty}' \sum_{l=1}^\infty (2l+1)\left[\frac{1}{ 1-\lambda_l^{TM}}\frac{\partial \lambda_l^{TM}}{\partial \beta}+
\frac{1}{1-\lambda_l^{TE}}\frac{\partial \lambda_l^{TE}}{\partial \beta}\right].
\end{equation}
Here the partial derivatives with respect to $\beta$ are most conveniently calculated on an analytic computer, on the basis of the expressions (2)-(6). If  series approximation for the eigenvalues $\lambda_l^{TM}$ or $\lambda_l^{TE}$ were accessible, for instance in either of the temperature limits, it would be convenient to use Eq.~(27) for evaluating approximate expressions for $E$. Obviously, $E\rightarrow 0$ when $d\rightarrow \infty$.

5.   It should be noted that since the spherical two-surface geometry that we are considering in our paper is different from the conventional parallel-plates geometry, this becomes reflected in the way in which we define the nondimensional temperature: we define it as $t=2\pi a/\beta$, i.e., with the inner radius $a$ as the geometrical variable instead of the conventional gap distance $d$. This is a natural definition in the case of curved surfaces. There are thus {\it two} different temperature scales involved here. This implies that at ordinary room temperature our problem becomes a {\it high-temperature} problem: by taking $T=300$ K, $a=1$ mm we obtain $t$ to be as large as about 830. Under these circumstances, it follows from our Figs. 6 and 7 that the $m=0$ TE term would be most important. A measurement of the surface surface force in this case would thus be critical. So far, no measurement of this force exists, however. So far, to our knowledge no conflict between our theory and experiments has been found.

6. Recently, it has been argued by Klimchitskaya and Mostepanenko \cite{klimchitskaya01} and Bordag {\it et al.} \cite{bordag00} that the Drude dispersive model leads to inconsistencies at low frequencies, even in the conventional case of planar geometry. The reason for this, according to these authors, is that the Drude relation leads to a discontinuity in the reflection coefficient $r_2$ as the imaginary frequency $\hat{\omega}\rightarrow 0$, in the case of perpendicular polarization. The plasma dispersion relation, instead of the Drude relation, is accordingly given preference by these authors since this discontinuity is not found to be present if one uses the plasma relation.

These arguments are quite interesting, since they raise doubts not only about the validity of the Drude relation as such, but more generally even about the applicability of the Lifshitz formula at low temperatures. We intend to return to a study of this problem in a later paper \cite{hoye02}. The problem with $\hat{\omega} \rightarrow 0$ is most naturally discussed if one assumes planar geometry from the outset. Here, let us merely make a few remarks, related to our treatment above, choosing a specific metal for concreteness. When the radii $a$ and $b$ are large, our spherical system of course approaches that of planar geometry. Let us take aluminum, for which one has \cite{klimchitskaya01,palik98} 
\[ \omega_p=1.9\times 10^{16}\, s^{-1}, \]
\begin{equation}
\gamma=9.6\times 10^{13}\, s^{-1}.
\end{equation}
\label{28}
For parallel plates separated by a gap $d$, the Matsubara frequencies (in dimensional units) are $\hat{\omega}_m=2\pi k_BTm/\hbar$. Zero-temperature theory is applicable as long as $dk_BT/\hbar c \ll 1$. At $T=0$, the contribution from $m=0$ is negligible, since this contribution is completely buried in the Matsubara frequency integral.

Assume now room temperature, $T=300$ K. Then for aluminum
\begin{equation}
\hat{\omega}_m=(2.48 m)\times 10^{14}\, s^{-1},
\end{equation}
\label{29}
which shows that the difference between two adjacent Matsubara frequencies is in this case quite appreciable. The most important frequencies for the Casimir effect occur when $\hat{\omega}_m \sim 2\pi c/d$, corresponding to the ordinary frequency $\hat{\omega}_m/2\pi$ being of the order of the inverse transit time for photons between the two surfaces. This corresponds to $m\sim \hbar c/(dk_BT)$. Taking for definiteness the gap to be  $d=0.5\, \mu$m, we obtain $m\sim 15$ to be the most significant Matsubara numbers. This is so far separated from $m=0$ that one should without any further calculation expect the  contribution from  $m=0$  to be quite small.  And this agrees with the about 1 \% level of temperature correction following from a more detailed calculation \cite{klimchitskaya01}.

It is instructive to calculate also the conductivities, and the refractive indices, as following from the Drude model for the two lowest frequencies. For convenience we now use SI units. We first write the square of the refractive index, $n^2=\varepsilon/\varepsilon_0$, in the same form as in conventional low-frequency theory for metals:
\begin{equation}
n^2(i\hat{\omega})=1+\frac{\sigma(i\hat{\omega})}{\varepsilon_0\hat{\omega}}.
\end{equation}
\label{30}
Here $\sigma(i\hat{\omega})$ is an effective frequency-dependent conductivity. The Drude model, Eq.~(24), corresponds to
\begin{equation}
\sigma(i\hat{\omega})=\frac{\varepsilon_0\omega_p^2}{\hat{\omega}+\gamma}.
\end{equation}
\label{31}
For the static conductivity, using Eq.~(1), we find
\begin{equation}
\sigma(0)=3.33\times 10^7 \, {\rm S/m},
\end{equation}
\label{32}
whereas for the $m=1$ case
\begin{equation}
\sigma(i\hat{\omega}_1)=0.93 \times 10^7 \, {\rm S/m}.
\end{equation}
\label{33}
The effective conductivity thus diminishes quickly when we move away from $m=0$. The corresponding square refractive indices are
\begin{equation}
n^2(i\hat{\omega} \rightarrow 0)=(3.76/\hat{\omega})\times 10^{18}, 
\end{equation}
\label{34}
\begin{equation}
n^2(i\hat{\omega}_1)=(1.05/\hat{\omega_1})\times 10^{18}.
\end{equation}
\label{35}
These refractive indices are large. However, the important point is that when $\hat{\omega}\rightarrow 0$, Eq.~(34) shows explicitly how $n(i\hat{\omega})\hat{\omega}\rightarrow 0$ when $\hat{\omega}\rightarrow 0$. Thus, this approach is in agreement with our option B above, in Sec.~3.

Let us return to the reflection coefficient $r_2$ mentioned above, for perpendicular polarization. It is defined as \cite{klimchitskaya01}
\begin{equation}
r_2= \frac{p-s}{p+s}, 
\end{equation}
\label{36}
 where $s$ and $p$ are the conventional Lifshitz variables for planar geometry:
\[ s=(\varepsilon-1+p^2)^{1/2}, \]
\begin{equation}
k_\perp \equiv |{\bf k}_\perp |=(\hat{\omega} /c)(p^2-1)^{1/2}.
\end{equation}
\label{37}
The important question is: does $r_2$ really becomes discontinuous at $\hat{\omega}=0$ if one uses the Drude model? In our opinion, it does not. This can be seen from a power expansion in $\hat{\omega}/\gamma$ of the expressions above (we keep $k_\perp$ fixed; any normal metal must have a finite relaxation frequency $\gamma $). To lowest order we obtain $s-p \rightarrow \omega_p^2/(2\gamma k_\perp c)$, $s+p\rightarrow 2k_\perp c/\hat{\omega}$, resulting in 
\begin{equation}
r_2\rightarrow \frac{\omega_p^2}{4k_\perp^2 c^2}\frac{\hat{\omega}}{ \gamma}.
\end{equation}
\label{38}
This shows that $r_2$ goes to zero smoothly (in our case linearly) as $\hat{\omega}/\gamma \rightarrow 0$; no singularity at $\hat{\omega}=0$ is found.

We intend to discuss these points in more detail in the mentioned forthcoming paper \cite{hoye02}. There, we will also discuss the recent claim of Fischbach {\it et al.} \cite{fischbach01} that the results of Bostr\"{o}m and Sernelius \cite{bostrom00} come into conflict with experiment as well as with basic thermodynamics.

7.  Generally, when comparing the outcome of Casimir calculations with experiments, care should be taken if the calculation involves summation  over infinite series. It should here be observed that our discussion on the $m=0$ TE term in Section 3, as well as in our previous paper \cite{brevik02}, was based upon {\it statistical} methods, thus not involving summation methods like the one used by 
 van Kampen {\it et al.} \cite{kampen68} and others. Our viewpoint is quite physical: the static mode has to occur only {\it once}, not twice; it corresponds to the electric field being directed radially, thus transversely to the two spherical surfaces. This is precisely the TM mode.

8.  A remark on the so-called proximity force hypothesis \cite{blocki77} is in order, as this hypothesis is being made use of in connection with Casimir calculations for test bodies having spherical segments. Some doubts have been expressed in the literature concerning the accuracy of this hypothesis. The issue has recently been analyzed by Barton (personal communication), with the result that this hypothesis remains valid in fourth-order as well as in second-order perturbation theory. Actually the hypothesis holds to all orders, as was shown by Langbein \cite{langbein71}.

9.  Finally, the recent experiment of Bressi {\it et al.} \cite{bressi02} is interesting, since it reports a measurement of the Casimir force between conducting surfaces in a {\it parallel} configuration.  At present, an accuracy at a 15 \% level is achieved. It is to be hoped that this accuracy can be improved, although a direct experiment of this sort is obviously quite demanding. Again, as this is a low-temperature experiment, we cannot at present decide whether it is in accordance with our theory or not.

\bigskip

{\bf Acknowledgments}

\bigskip

I. B. thanks  Gabriel Barton, Vladimir Mostepanenko, and Roberto Onofrio for valuable correspondence.

\newpage

{\bf Figure Captions}

\bigskip
\bigskip

{\bf Figure 1}
Sketch of the spherical geometry.

\bigskip

{\bf Figure 2}
Logarithm of mutual nondimensional free energy, $\log_{10}(-\beta Ft)$, versus relative width $d/a$ for various values of the nondimensional temperature $t=2\pi a/\beta$. Refractive index $n=1.1$. 

\bigskip

{\bf Figure 3}
Logarithm $\log_{10}(-\beta F)$ versus $d/a$, when $n=1.1.$

\bigskip
 {\bf Figure 4}
Variation of $\log_{10}(-\beta Ft)$ versus $\log_{10}t$ for various values of $d/a$, when $n=1.1.$

\bigskip
{\bf Figure 5}
Same as Fig.~4, but with $n=2.0.$

\bigskip

{\bf Figure 6}
Same as Fig.~5, but for an ideal metal ($n=\infty$, the $m=0$ mode counted twice). Reproduced (with corrected labeling) from Ref.~\cite{hoye01}.

\bigskip

{\bf Figure 7}
Relative importance of the zero frequency term $m=0$ in the free energy, versus $t=2\pi a/\beta$. The quantity $Y$ is defined in Eq.~(26).

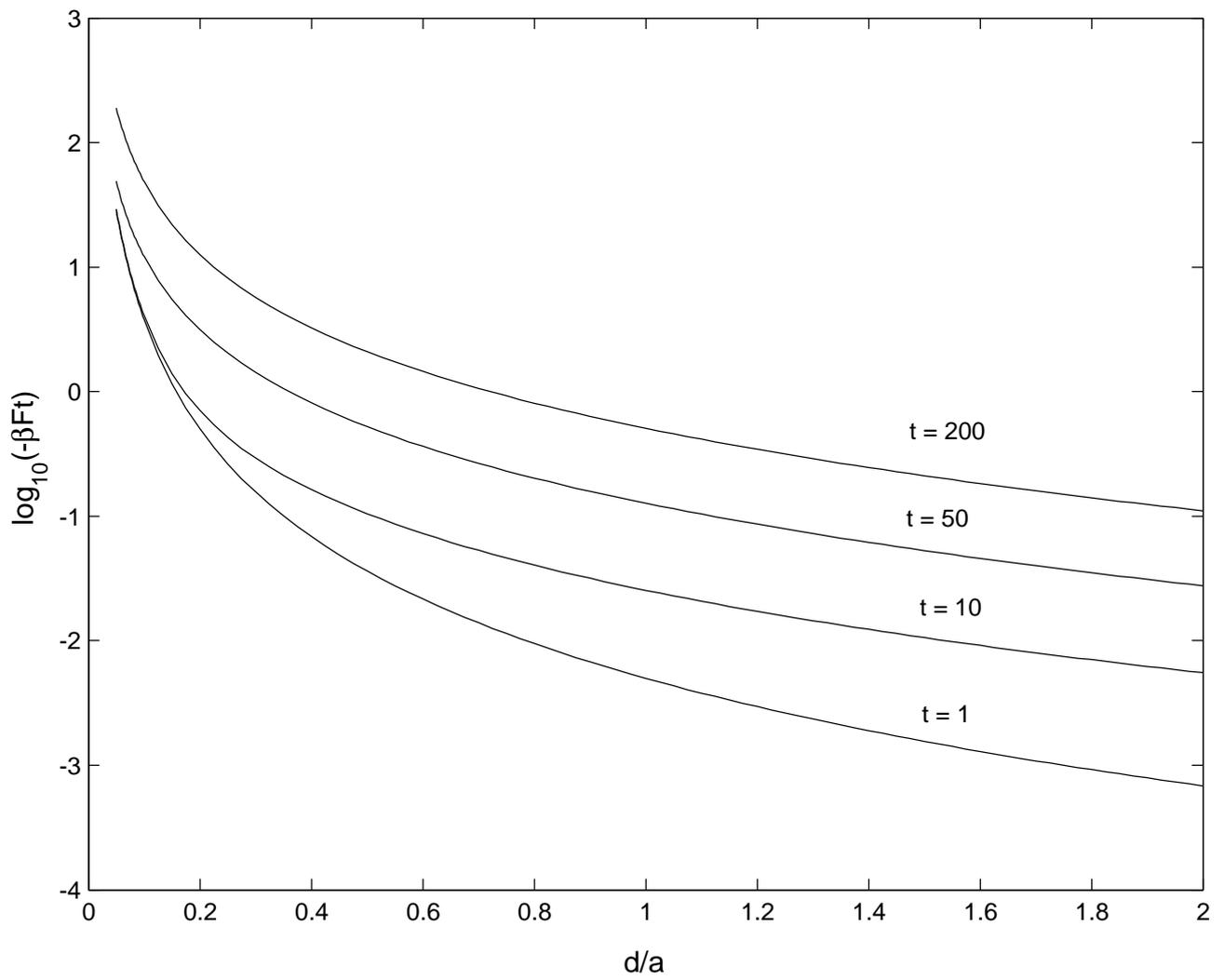
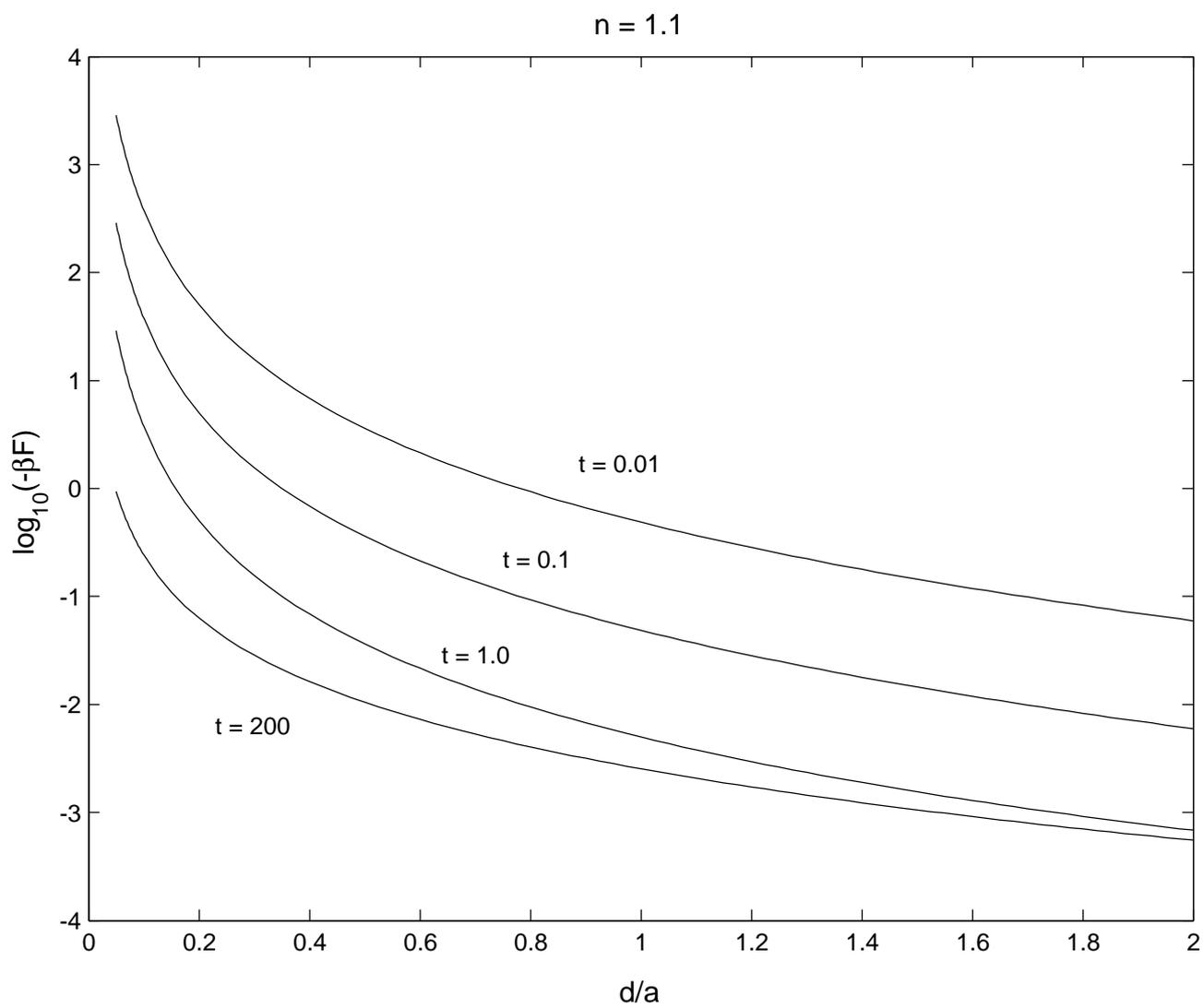
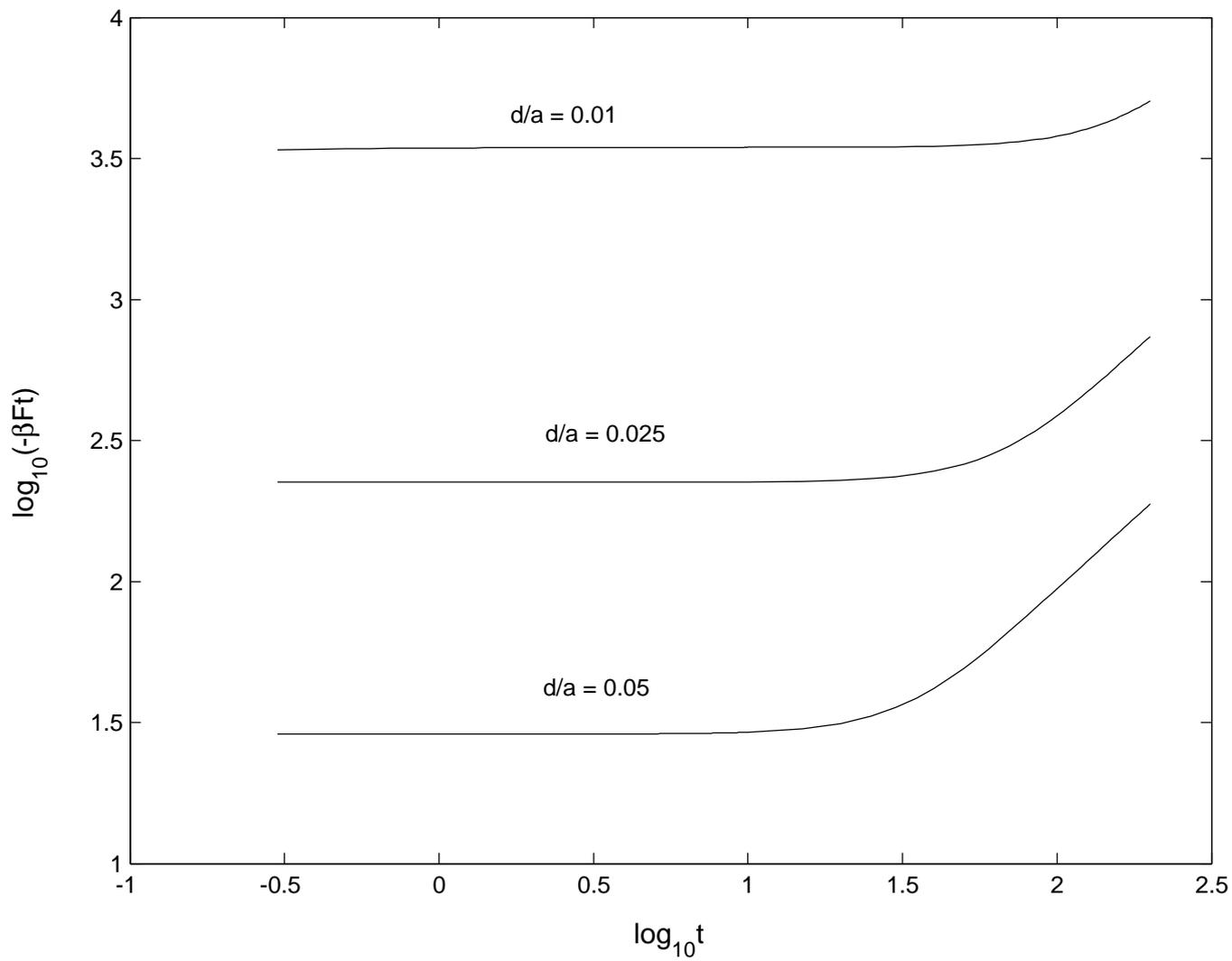
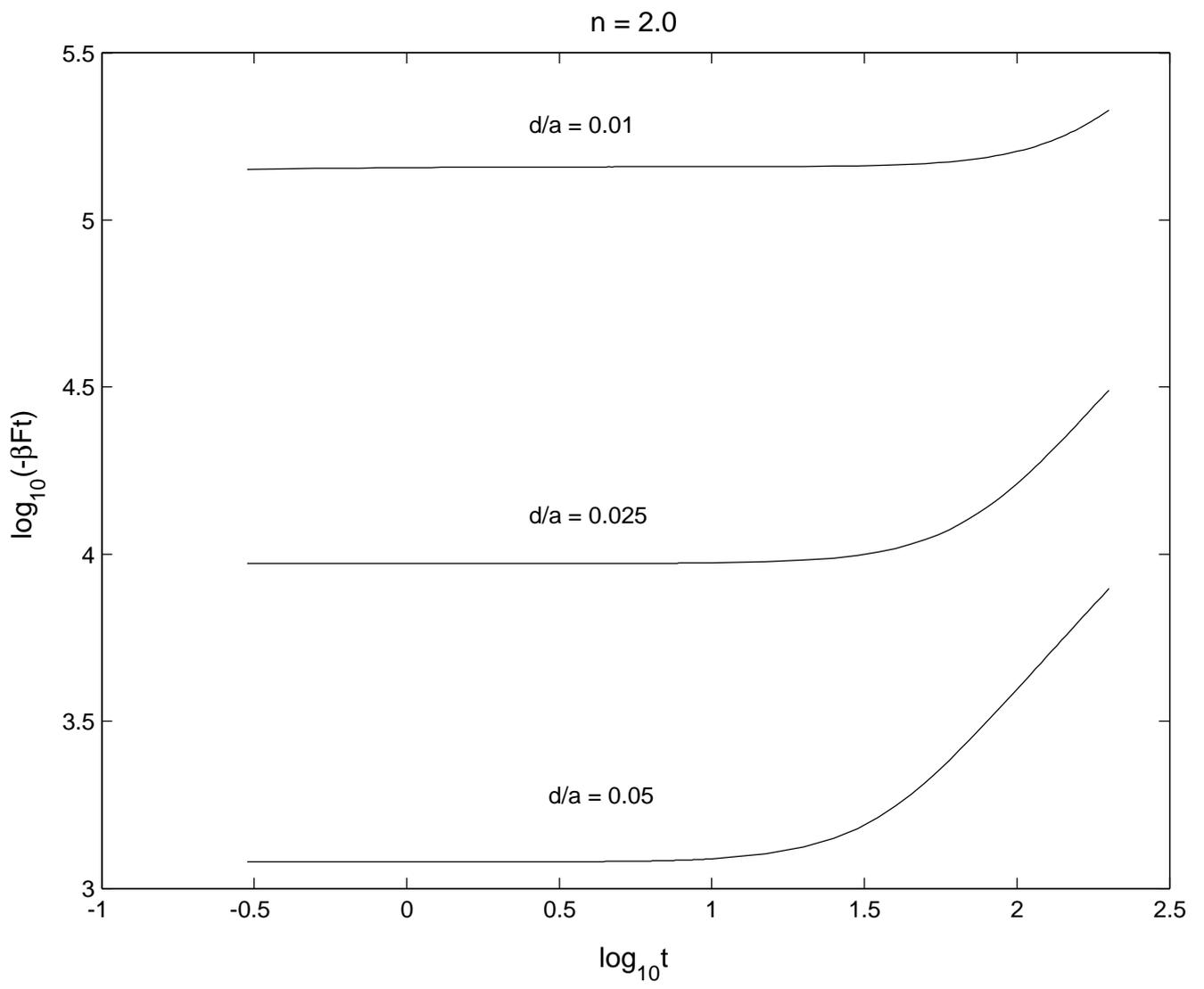
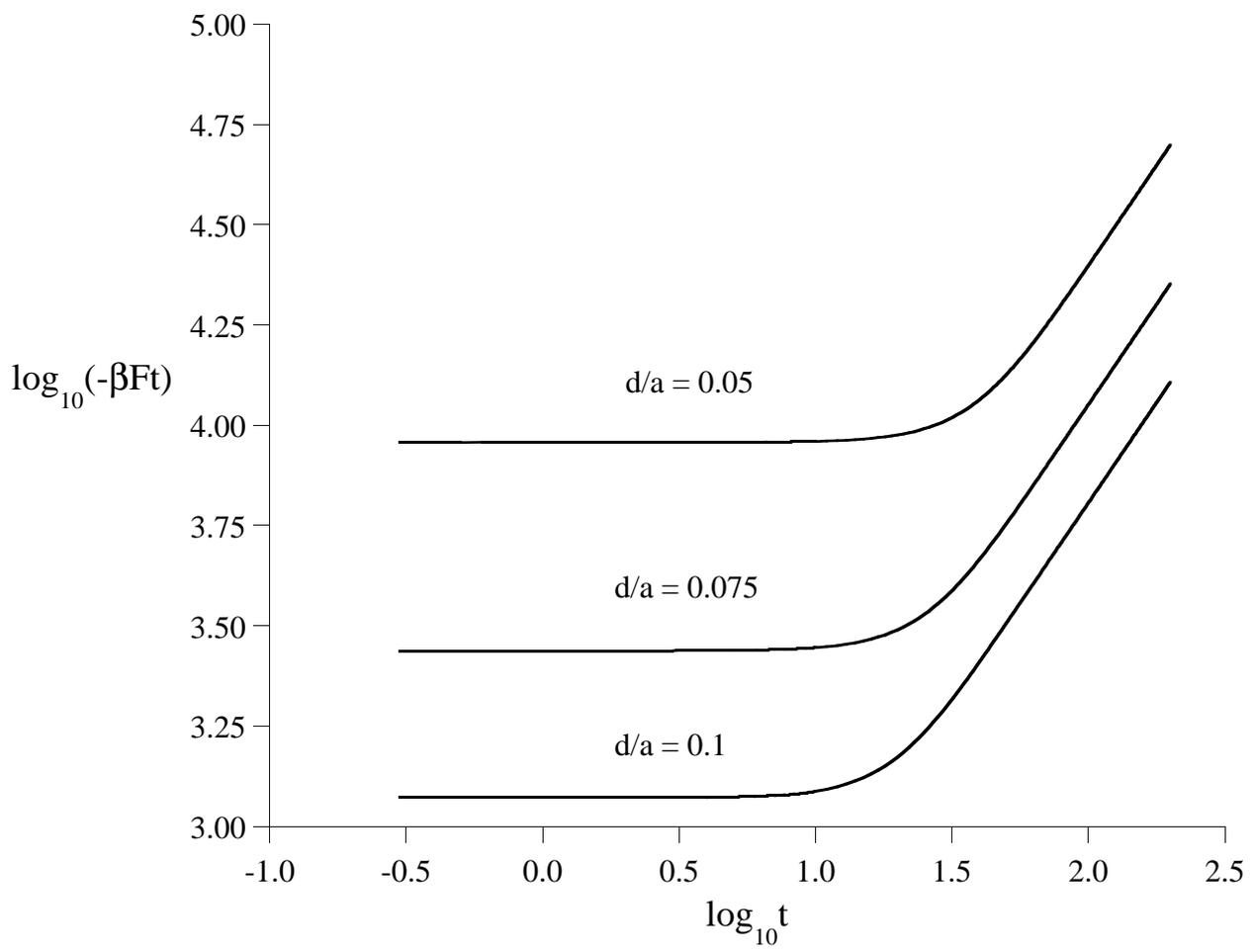
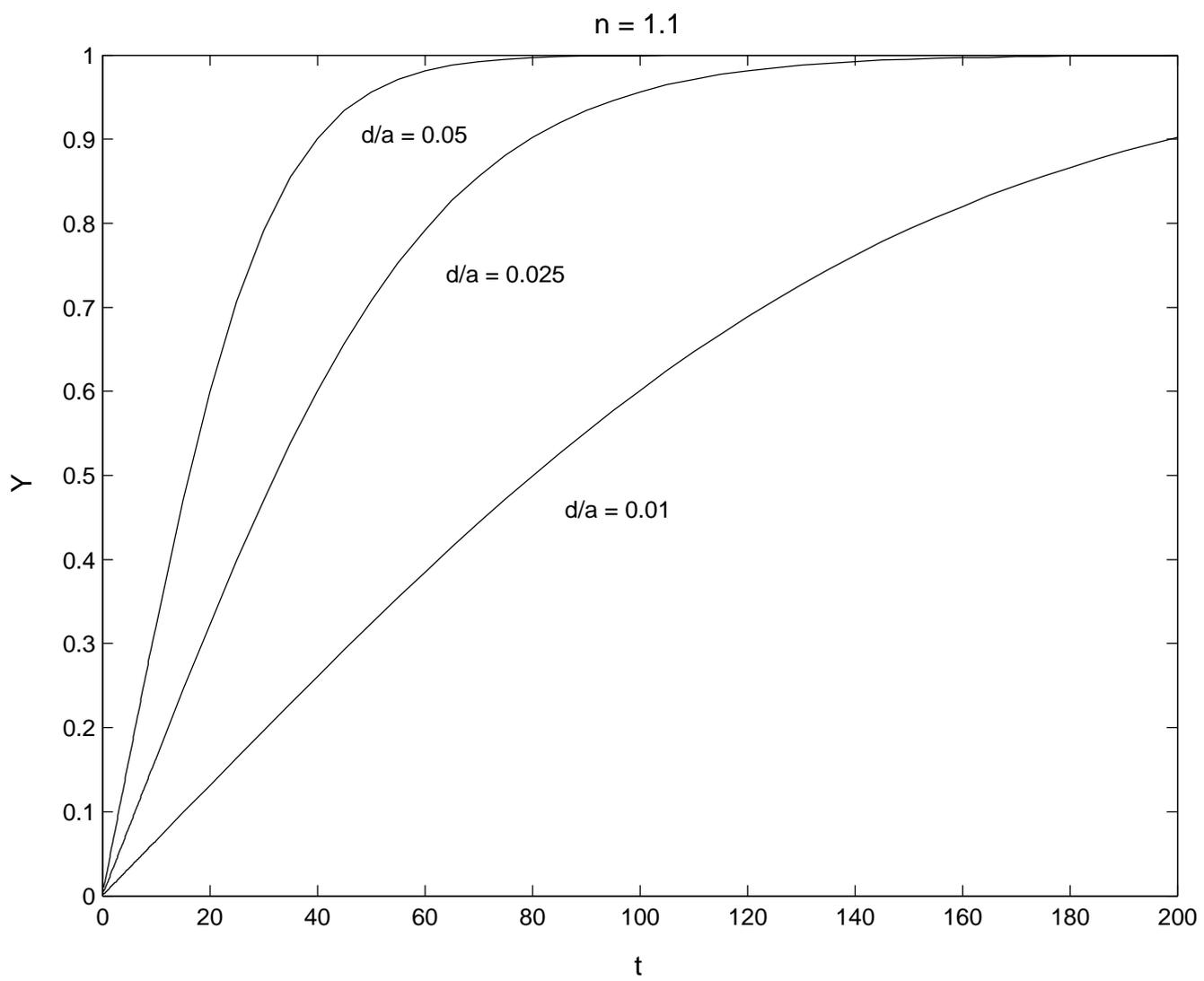
\end{document}